\documentclass{article} 
\usepackage{iclr2027_conference,times}

\usepackage{amsmath,amsfonts,bm}

\def\eqref#1{equation~\ref{#1}}

\def\floor#1{\lfloor #1 \rfloor}
\def\1{\bm{1}}

\DeclareMathAlphabet{\mathsfit}{\encodingdefault}{\sfdefault}{m}{sl}
\SetMathAlphabet{\mathsfit}{bold}{\encodingdefault}{\sfdefault}{bx}{n}

\usepackage{hyperref}
\usepackage{url}

\title{\ours: Pairwise Distortion-Free Watermarking Beyond Single-Token Entropy}

\author{Ruibo Chen$^{1,2,}$\thanks{This work was done while Ruibo Chen was interning at TikTok.}~,~Zhengmian Hu$^1$, ~Donghang Lu$^2$, ~Xuehao Cui$^1$,~Georgios Milis$^1$, \\
\textbf{Yihan Wu}$^1$, \textbf{Jian Du}$^{2,}$\thanks{Corresponding Authors}~,~\textbf{Heng Huang}$^{1,\dagger}$ \\
$^1$University of Maryland, College Park\quad $^2$TikTok \\
}

\usepackage{booktabs}
 \usepackage{multirow}
\newcommand{\ours}{\textsc{TTMark}}
\usepackage{amsmath}
\usepackage{amssymb}
\usepackage{mathtools}
\usepackage{amsthm}

\usepackage{algorithm}     
\usepackage{algpseudocode}  
\usepackage{amsmath}  
\usepackage{amssymb}
\usepackage{booktabs}
\usepackage{multirow}
\usepackage{graphicx}
\usepackage{subcaption} 
\usepackage[normalem]{ulem}
\useunder{\uline}{\ul}{}
\usepackage{xcolor}
\usepackage{amsmath,amsthm}
\usepackage{mathtools}
\usepackage{bm}
\usepackage{float}
\usepackage{tabularx}
\usepackage{seqsplit}

\theoremstyle{plain}
\newtheorem{theorem}{Theorem}[section]

\theoremstyle{definition}

\theoremstyle{remark}

\usepackage{graphicx}    
\usepackage{subcaption}  

\iclrfinalcopy 
\begin{document}

\maketitle

\begin{abstract}
Distortion-free watermarking enables reliable attribution of machine-generated text while preserving output distribution. However, existing methods operate independently on each generated token, making their detection capability fundamentally constrained by the entropy of the next-token distribution. We present \textbf{T}andem \textbf{T}oken Water\textbf{Mark} (\ours), a general pairwise watermarking framework that extends distortion-free watermarking from individual tokens to adjacent token pairs. By watermarking the joint distribution of consecutive tokens, \ours\ enlarges the effective watermarking alphabet from $V$ to $V^2$, allowing the detector to exploit both token entropy and conditional entropy while preserving distortion-freeness over the joint distribution. We further introduce a branch-isolating concatenated tandem generation algorithm that efficiently constructs the joint distribution in a single forward pass. Theoretically, we show that pairwise watermarking achieves better expected detection strength in low-entropy regimes. Extensive experiments across multiple language models, datasets, and three representative distortion-free watermarking schemes demonstrate that \ours\ consistently improves detectability without degrading generation quality, while also improving robustness to edits and enhancing localized watermark detection.

\end{abstract}

\section{Introduction}

The rapid deployment of large language models has necessitated the development of reliable and minimally invasive techniques for identifying machine-generated text. This imperative is further underscored by emerging regulatory frameworks, such as the EU AI Act~\citep{eu-ai-act}, which mandate transparency and traceability for AI-generated content. Watermarking~\citep{Aaronson2022,kirchenbauer2023watermark,zhao2023provable} addresses this challenge by embedding a statistical signal during the decoding process, enabling a detector equipped with a secret key to reliably distinguish watermarked outputs from human-written text. Among existing paradigms, distortion-free watermarking~\citep{hu2023unbiased,dathathri2024scalable,chen2025improved} is particularly compelling: when marginalized over the secret key, the watermarked decoding distribution remains strictly identical to the original model distribution, thereby precluding the introduction of systematic distributional bias.

Despite these advantages, watermark detectability degrades significantly in practical deployment scenarios. Modern decoding pipelines heavily rely on temperature scaling, top-$p$ sampling, and top-$k$ truncation to improve output quality, which simultaneously reduces the entropy of the next-token distribution. Because standard watermarking schemes operate token-by-token, their statistical power is strictly limited by the available uncertainty at each step~\citep{huang2023towards,dathathri2024scalable,chen2026more}. This constraint becomes critical during short generations, under conservative sampling, or in localized detection tasks, where the scarcity of high-entropy tokens severely diminishes the detector's observation pool.

To overcome this bottleneck, we propose \textbf{T}andem \textbf{T}oken Water\textbf{Mark} (\ours), a novel watermarking mechanism that shifts the domain of distortion-free watermarking from individual tokens to adjacent token pairs. By reweighting the joint distribution of consecutive tokens rather than the standard next-token distribution, \ours\ expands the effective vocabulary size from $V$ to $V^2$. This formulation naturally surfaces both the entropy of the immediate token and the conditional entropy of its successor. Crucially, \ours\ maintains the distortion-free property over the joint space, ensuring that the original token marginals are perfectly preserved in expectation. Naively constructing the joint distribution requires independent evaluations for every candidate first token. \ours\ circumvents this by proposing \textbf{branch-isolating concatenated tandem generation}. By concatenating the first tokens together with a branch-isolating attention mask, all candidate continuations can be evaluated in a single forward pass, producing the required conditional distributions. For evaluation, we provide a direct detector for unedited text alongside an alignment-based detector robust to perturbations. 

Theoretically, we show that in low-entropy regimes, token pairing yields an approximate $\sqrt{2}$ improvement in the aggregate $z$-score. More broadly, we prove the expected $z$-score of our distortion-free pairwise watermark is lower-bounded by the single-token baseline. Therefore, \ours\ can further exploit distributional uncertainty that single-token watermarking leaves unutilized. Empirically, we demonstrate that \ours\ universally improves baseline distortion-free watermarking schemes without degrading output quality. When evaluated on the C4 dataset~\citep{raffel2020exploring} with Llama-3.2-3B-Instruct at a 0.001\% false positive rate, the tandem approach elevates the true positive rate of SynthID-Text~\citep{dathathri2024scalable} from 50.3\% to 73.2\%, and ENS-MCMark~\citep{wu2025ensemble} from 54.0\% to 81.8\%. We confirm that these gains generalize consistently across different datasets, model families, and stringent sampling conditions. Moreover, the proposed pairwise watermark enhances robustness to perturbations and substantially reduces false negatives in localized detection contexts, while maintaining generation quality.

In summary, our primary contributions are as follows:

\vspace{-2mm}
\begin{itemize}
    \item We introduce \ours, a pairwise watermarking mechanism that preserves the distortion-free property while expanding the watermarking alphabet from $V$ to $V^2$. To operationalize this efficiently, we develop a branch-isolating concatenated tandem generation procedure alongside robust direct and alignment-based detectors.
    \item We theoretically establish that operating over joint token distributions exploits residual conditional entropy, granting access to stronger detection signals than single-token baselines.
    \item We present extensive empirical evidence demonstrating that \ours\ significantly improves detectability and robustness across diverse language models, existing watermarking schemes, low-entropy sampling regimes, and semantic-preserving attacks.
    \vspace{-1mm}
\end{itemize}

\section{Related Work}

\vspace{-1mm}
\subsection{Language Model Watermarking}
\vspace{-1mm}
The rapid advancement of large language models (LLMs) has raised concerns about content authenticity and potential misuse~\citep{grinbaum2022ethical, crothers2023machine, pei2026deepfake, yang2024survey}.
To address these concerns, \citet{Aaronson2022} proposed an unbiased watermarking framework based on the Gumbel-max trick, using prefix $n$-grams as watermark keys to guide pseudo-random token selection.
\citet{christ2023undetectable} extended inverse-sampling-based watermarking to binary language models using position-dependent keys.
ITS-edit and EXP-edit~\citep{kuditipudi2023robust} further refined these approaches by introducing predefined watermark key sets.
\citet{hu2023unbiased} developed an inverse-sampling method with model-driven log-likelihood ratio (LLR) detection, as well as $\gamma$-reweight, a logits-based unbiased watermarking scheme.
STA-1~\citep{mao2024watermark} improved generation quality in low-entropy regimes through rejection sampling, while SynthID~\citep{dathathri2024scalable} employed tournament sampling to enhance detectability.
Building on $\gamma$-reweight, DiPmark~\citep{wu2023dipmark} introduced model-agnostic detection, and MCmark~\citep{chen2025improved} further improved the detectability of unbiased watermarking.

\subsection{Joint and Multi-Token Language Modeling}
\vspace{-1mm}
Recent approaches improve inference efficiency by predicting multiple future tokens rather than generating them strictly autoregressively.
Block-wise decoding uses parallel prediction heads to reduce sequential dependencies~\citep{stern2018blockwise}.
Speculative decoding adopts a draft-and-verify paradigm, in which a lightweight model proposes multiple tokens for parallel verification by a larger model while preserving its output distribution~\citep{leviathan2023fast}.
Subsequent work improves sampling efficiency~\citep{chen2023accelerating} and introduces tree-based verification to evaluate multiple speculative branches simultaneously~\citep{miao2024specinfer}.

\section{Method}
\subsection{Preliminary: Distortion-Free Watermarking}

Let \( V \) denote the vocabulary, with cardinality \( N = |V| \).
Given a prompt \(p\), a language model \( M \) generates tokens autoregressively.
At time step \( t \), conditioned on the preceding sequence \( \bm{x}_{1:t} \) and prompt \(p\), the probability of generating the next token \( x_{t+1} \in V \) is
\( P_M(x_{t+1} \mid \bm{x}_{1:t}, p) \).
Watermarking reweights the original distribution \( P_M(\cdot \mid \bm{x}_{1:t}, p) \) into a watermarked distribution
\( P_w(\cdot \mid \bm{x}_{1:t}, p, k) \), where \( k \in \mathcal{K} \) is a private watermark key sampled from a known distribution
\( P_{\mathcal{K}} \) over the key space \( \mathcal{K} \).
Following prior work~\citep{Aaronson2022,wu2023dipmark,dathathri2024scalable}, a watermarking algorithm is \emph{distortion-free} if it preserves the original token distribution in expectation over the key space.
Formally, for any \( P_M \in \mathcal{P} \) and any token \( x_{t+1} \in V \),
\begin{equation}\label{eq:unbiased}
\mathbb{E}_{k \sim P_{\mathcal{K}}}
    \left[P_w(x_{t+1} \mid \bm{x}_{1:t}, p, k)\right]
    = P_M(x_{t+1} \mid \bm{x}_{1:t}, p).
\end{equation}

For detection, distortion-free watermarks typically define a scoring function \(s: V \times \mathcal{K} \rightarrow \mathbb{R}\), which assigns score \(s(x,k)\) to token \(x\) under key \(k\).
Under the null hypothesis that no watermark is present, the expected score for token \(x\) is \(\mathbb{E}_{k \sim P_{\mathcal{K}}} s(x,k)\).
For model-agnostic watermarks, where the original output logits are unavailable, this expectation is token-independent:
\(\mathbb{E}_{k \sim P_{\mathcal{K}}} s(x,k)=\mu\) for all \(x \in V\).
The score variance is likewise independent of \(x\), i.e., \(\text{Var}_k[s(x,k)] = \sigma^2\) for all \(x \in V\).

Detection then proceeds with a \(z\)-score test:
\begin{equation}
    z = \frac{\sum_{t=1}^T s(x_t,k_t) - \mu T}{\sqrt{T}\sigma},
\end{equation}

where \(k_t\) is the watermark key associated with token \(x_t\).
If the resulting \(z\)-score exceeds a predetermined threshold, the sequence is classified as watermarked.

\subsection{Tandem Token Watermark}\label{sec:ttmark}

\begin{figure}
    \centering
    \includegraphics[width=0.95\linewidth]{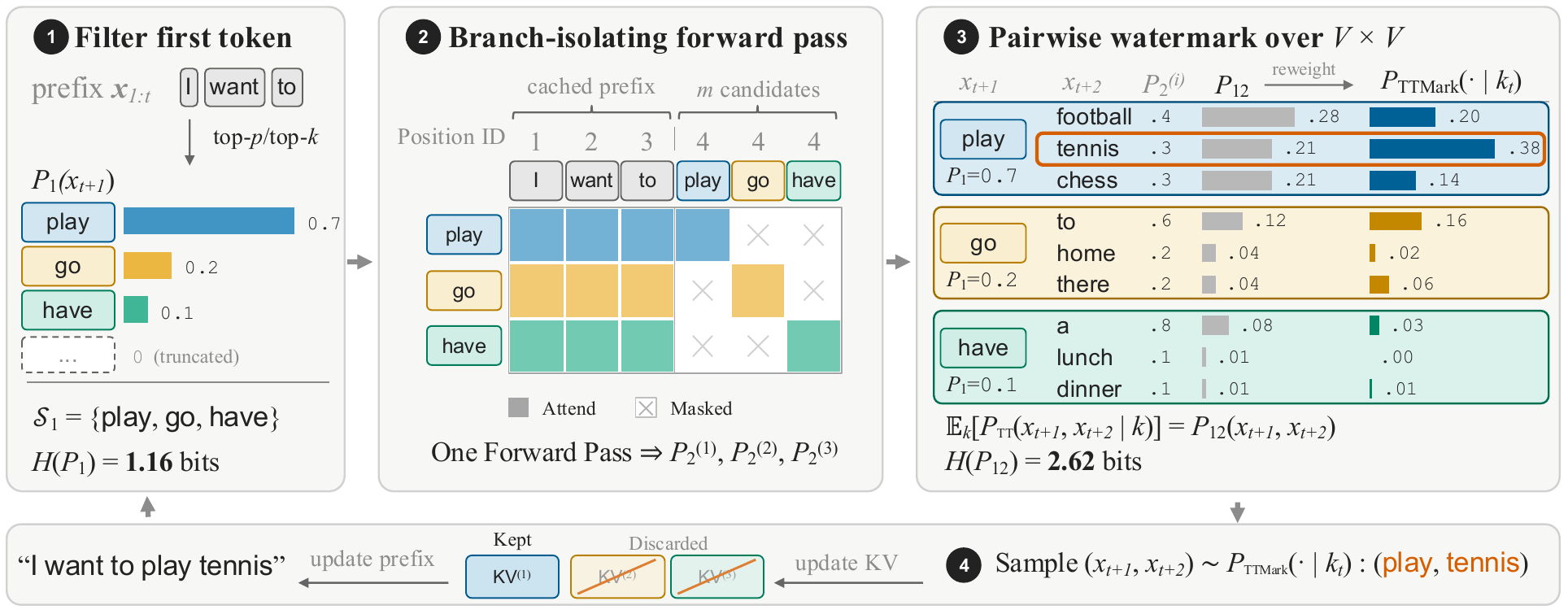}
    \vspace{-1mm}
    \caption{Overview of branch-isolating concatenated tandem generation. \ours\ first forms a filtered set of candidate next tokens and evaluates their continuations in a single batched forward pass using a branch-isolating attention mask. The resulting conditional distributions are combined with the first-token probabilities to construct a joint two-token distribution, enabling efficient pairwise watermarking over adjacent tokens.}
    \vspace{-3mm}
    \label{fig:pipeline}
\end{figure}

We propose \textbf{T}andem \textbf{T}oken Water\textbf{Mark} (\ours), which applies watermarking to a two-token block rather than to each token independently. The full pipeline is illustrated in Fig.~\ref{fig:pipeline}.
At time step \(t\), \ours\ operates on the joint next-token distribution
\(P_M(x_{t+1},x_{t+2}\mid \bm{x}_{1:t},p)\).
By autoregressive factorization, this distribution can be written as
\begin{equation}\label{eq:joint-next-token-factorization}
    P_M(x_{t+1},x_{t+2}\mid \bm{x}_{1:t},p)
    =
    P_M(x_{t+1}\mid \bm{x}_{1:t},p)
    P_M(x_{t+2}\mid \bm{x}_{1:t},x_{t+1},p).
\end{equation}
We describe an efficient procedure for constructing this joint distribution in Sec.~\ref{sec:generation}. \ours\ then reweights the joint distribution into
\(P_\mathrm{TTMark}(x_{t+1},x_{t+2} \mid \bm{x}_{1:t},p,k)\), while preserving distortion-freeness at the level of token pairs:
\vspace{-0.5mm}
\begin{equation}
    \mathbb{E}_{k\sim P_{\mathcal{K}}}
    \left[
    P_\mathrm{TTMark}(x_{t+1},x_{t+2} \mid \bm{x}_{1:t},p,k)
    \right]
    =
    P_M(x_{t+1},x_{t+2} \mid \bm{x}_{1:t},p).
\end{equation}

The detector is modified accordingly: instead of scoring a single token, it uses a pairwise scoring function
\(s: V^2\times\mathcal{K}\rightarrow \mathbb{R}\) and assigns score \(s((x_{t+1},x_{t+2}),k)\) to each generated token pair.
Intuitively, \ours\ enlarges the effective vocabulary size from \(V\) to \(V\times V\), increasing the amount of distributional uncertainty available for watermarking.
We next make this intuition precise from two perspectives.

\paragraph{Entropy.}
Entropy is a primary bottleneck for watermark signal strength, as it limits the expected per-step \(z\)-score.
To illustrate the effect of two-token watermarking, consider a setting in which each one-token conditional distribution has an average entropy of \(h\). In the low-entropy regime, following prior work~\citep{huang2023towards,dathathri2024scalable,chen2026more}, the expected watermark strength, represented by the expected single-step \(z\)-score, can be approximated as scaling linearly with entropy, i.e., \(z_t \propto ah\) for some constant \(a\).
For a single-token watermark over \(T\) tokens, the aggregate \(z\)-score scales as $z \propto \frac{Tah}{\sqrt{T}}=\sqrt{T}ah$. If watermarking is instead applied to \(T/2\) token pairs, then for each pair, the joint entropy is approximately \(2h\). The corresponding aggregate score scales as:
\vspace{-1mm}
\begin{equation}
    z_\mathrm{TTMark} \propto \frac{(T/2)a(2h)}{\sqrt{T/2}}=\sqrt{2T}ah,
\end{equation}
which suggests a \(\sqrt{2}\) gain under this approximation.

\paragraph{Expected \(z\)-Score Upper Bound.}
We further show that, for any distortion-free watermark, grouping two tokens yields no smaller upper bound on the expected normalized \(z\)-score.
For a single-token watermark, let
\(z(x,k)=(s(x,k)-\mu)/\sigma\) denote the normalized score for token \(x\).

\begin{theorem}\label{thm:single-step-z-bound}
Let \(P\) be the unwatermarked distribution over vocabulary \(V\), and let \(P_w(\cdot\mid k)\) be any distortion-free watermark satisfying
\(\mathbb{E}_{k\sim P_{\mathcal{K}}}\left[P_w(x\mid k)\right]=P(x)\) for all \(x\in V\).
Assume that \(\mathbb{E}_{k}\left[z(x,k)\right]=0\) and \(\operatorname{Var}_{k}[z(x,k)]=1\) for every \(x\).
Then
\begin{equation}
    \mathbb{E}_{k\sim P_{\mathcal{K}},\,x\sim P_w(\cdot\mid k)}
    \left[z(x,k)\right]
    \leq
    \sum_{x\in V} \sqrt{P(x)-P^2(x)}.
\end{equation}
\end{theorem}
\vspace{-0.5mm}
The proof is deferred to Appendix~\ref{app:proof-single-step-z-bound}.

\begin{theorem}\label{thm:pairwise-z-bound}
Let \(f(Q)=\sum_{x}\sqrt{Q(x)-Q^2(x)}\) denote the upper bound from the preceding theorem.
For any joint distribution \(P_{12}\) on \(V\times V\), with marginal \(P_1\) and conditional distributions \(P_{2\mid 1=x_1}\), the pairwise upper bound satisfies
\vspace{-0.5mm}
    \begin{equation}
        f(P_{12})
        \ge
        \frac{
        f(P_1)+
        \mathbb{E}_{x_1\sim P_1} f(P_{2\mid 1=x_1})
        }{\sqrt{2}}.
    \end{equation}
    \vspace{-0.5mm}
\end{theorem}
The factor \(\sqrt{2}\) arises from the normalization when two token-level contributions are combined into one statistic. The proof is deferred to Appendix~\ref{app:proof-pairwise-z-bound}. Theorem~\ref{thm:pairwise-z-bound} provides a justification for watermarking the joint two-token distribution. It shows that for any distortion-free watermark, the pairwise upper bound is at least as large as the normalized aggregate of the single-token bound and the expected conditional bound for the second token.

\vspace{-1mm}
\subsection{Branch-Isolating Concatenated Tandem Generation}\label{sec:generation}
\vspace{-1mm}

Directly constructing the pair distribution required by \ours\ is computationally expensive, since the conditional distribution of \(x_{t+2}\) depends on the realized value of \(x_{t+1}\).
Under standard decoding, however, the next-token logits are transformed by temperature scaling and truncation rules such as Top-P or Top-K sampling, which produce a sparse decoding distribution.
Let \(P_1\) denote the filtered distribution for \(x_{t+1}\), and let
\(\mathcal{S}_1=\{x_{t+1}^{(i)}\}_{i=1}^{m}\) be its nonzero support, where typically \(m\ll |V|\).
For each candidate \(x_{t+1}^{(i)}\in\mathcal{S}_1\), we must compute the filtered distribution of \(x_{t+2}\) conditioned on that candidate. To ensure the efficiency of the generation, we propose \emph{Branch-Isolating Concatenated Tandem Generation}, where the candidates are concatenated together, evaluated all in one forward pass, while preventing different candidate branches from attending to one another. The generation algorithm is illustrated in Alg.~\ref{alg:generation}.

Let the current prefix cache have length \(L\), and let the concatenated candidate block be
\(\bm{x}_{t+1}^{(1:m)}=[x_{t+1}^{(1)},\ldots,x_{t+1}^{(m)}]\).
The branch-isolating attention mask \(A\) allows the query associated with branch \(r\) to attend to all cached prefix positions and to the corresponding candidate token \(x_{t+1}^{(r)}\), while preventing attention to other candidate branches:
\vspace{-0.5mm}
\begin{equation}
    A_{r,j} =
    \begin{cases}
        1, & 1\le j\le L,\\
        1, & j=L+r,\\
        0, & \text{otherwise},
    \end{cases}
    \qquad r=1,\ldots,m.
    \label{eq:masked_generation}
    \vspace{-2mm}
\end{equation}

With this mask, the batched computation is equivalent to executing \(m\) independent one-token continuations from the same prefix cache.
To preserve the same positional encoding as independent decoding, all candidate branches are assigned position index \(L+1\).

The same decoding parameters are then applied to each branch-specific second-token distribution.
Let \(P_2^{(i)}\) be the filtered distribution for \(x_{t+2}\) after conditioning on \(x_{t+1}^{(i)}\), and let
\(\mathcal{S}_2^{(i)}=\{x\in V:P_2^{(i)}(x)>0\}\) denote its support.
The admissible pair support is therefore
\(\{(x_{t+1}^{(i)},x_{t+2}):x_{t+1}^{(i)}\in\mathcal{S}_1,\;x_{t+2}\in\mathcal{S}_2^{(i)}\}\).
Following the autoregressive factorization in Eq.~\ref{eq:joint-next-token-factorization}, the pre-watermark mass assigned to an admissible pair is
\(P_1(x_{t+1}^{(i)})P_2^{(i)}(x_{t+2})\).
This filtered pair distribution is reweighted into
\(P_\mathrm{TTMark}(x_{t+1},x_{t+2}\mid \bm{x}_{1:t},p,k)\), from which the next token pair is sampled.
After sampling, only the KV cache corresponding to the selected first-token branch is retained, and this cache is advanced with the sampled second token to initialize the next iteration.
Consequently, each two-token generation step requires one batched branch evaluation and one cache update along the selected branch, rather than \(m\) separate conditional forward passes.

\vspace{-1.5mm}
\paragraph{Space Efficiency Analysis.} This procedure keeps the memory overhead nearly the same. The algorithm only creates temporary KV states and logits for the \(m_t\) candidate tokens at the current step, and then discards all but the selected branch. Thus the persistent cache size is essentially the same as ordinary autoregressive decoding.

\vspace{-1.5mm}
\paragraph{Time Efficiency Analysis.}
From a purely serial computation perspective, if a batched step over \(m\) candidates were charged as \(m\) independent cached forward passes, producing two tokens would cost \(m+1\) cached steps instead of the two steps used by standard decoding, giving an approximate time-cost ratio of \((m+1)/2\). However, in actual GPU decoding, latency is often bounded by memory bandwidth because each cached step must read the model weights and KV states from memory. Branch-isolating generation evaluates the \(m\) candidate continuations in one batched pass, which amortizes these memory reads across branches. Therefore, the observed wall-clock overhead can be much smaller than the serial ratio \((m+1)/2\), which is reported in Sec.~\ref{sec:analysis-ablation}. Furthermore, \ours\ reduces computational overhead by applying the watermarking algorithm only once for every two tokens. It should be noted that watermarking algorithms can be highly computationally expensive, frequently necessitating large temporary matrix allocations and the strictly serial execution inherent to multi-layer reweighting architectures, such as SynthID-Text.

\vspace{-1mm}
\subsection{Detection}\label{sec:detection}
\vspace{-1mm}
\paragraph{Direct Detection.}
We first consider the setting in which the generated sequence is observed without token-level corruption.
Let \(\bm{y}_{1:L}\) denote the sequence to be tested and let \(n=\floor{L/2}\) be the number of complete token pairs.
\ours\ partitions the sequence into the same non-overlapping pairs \((y_{2i-1},y_{2i})\), for \(i=1,\ldots,n\), used during generation.
The final unpaired token, if \(L\) is odd, is discarded for detection.
Given the secret key stream \(\{k_t\}_{t=1}^{L}\), the detector evaluates the pairwise score \(s((y_{2i-1},y_{2i}),k_{2i-1})\).
Under the null hypothesis that the text is not watermarked, the key is independent of the observed text; hence the pairwise score has key-averaged mean \(\mu\) and variance \(\sigma^2\), analogously to the single-token detector.
The direct detection statistic is therefore
\begin{equation}
    z_{\mathrm{dir}}
    =
    \frac{
        \sum_{i=1}^{n} s((y_{2i-1},y_{2i}),k_{2i-1}) - n\mu
    }{
        \sqrt{n}\sigma
    } .
    \label{eq:direct-detection}
\end{equation}
For sufficiently long sequences, \(z_{\mathrm{dir}}\) is compared against a threshold calibrated from the standard normal approximation or from an empirical null distribution.
This procedure is the exact pair-vocabulary analogue of single-token watermark detection, with \(V\) replaced by \(V^2\).

\vspace{-1mm}
\paragraph{Empirical Detection.}
Direct detection assumes that the detector observes the same pair boundaries as the generator.
In practice, paraphrasing, insertion, deletion, or partial retokenization can shift these boundaries, causing otherwise watermarked pairs to be evaluated under the wrong alignment.
To make detection robust to such perturbations, we use a penalized alignment procedure that searches over subsequences of adjacent token pairs while discouraging excessive skipping.

Let \(\beta\ge 0\) be a skip penalty.
For an observed sequence \(\bm{y}_{1:L}\), we compute the best penalized score by dynamic programming over token positions.
Let \(D_t\) be the maximum score obtainable after reading the first \(t\) observed tokens.
The recurrence is
\begin{equation}
    D_t
    =
    \max\left\{
        D_{t-1}-\beta,\;
        D_{t-2}+z((y_{t-1},y_t),k_{t-1})
    \right\},
    \label{eq:empirical-detection-dp}
\end{equation}
where the first transition skips token \(y_t\), and the second transition retains the adjacent pair \((y_{t-1},y_t)\) using the positional key \(k_{t-1}\).
The boundary condition is \(D_0=0\), and the pair-retention transition is valid only when \(t\ge 2\).
The resulting statistic is
\begin{equation}
    z_{\mathrm{emp}}
    =
    \frac{D_L}{\sqrt{n}},
    \qquad n=\floor{L/2}.
    \label{eq:empirical-detection-score}
\end{equation}
We normalize by the original number of complete pairs \(n\), so that the detector cannot inflate the statistic by selecting only a small number of high-scoring pairs.
Because the maximization over alignments changes the null distribution of the score, thresholds for \(z_{\mathrm{emp}}\) are calibrated empirically on non-watermarked text generated or collected under the same evaluation protocol. The full algorithm is detailed in Alg.~\ref{alg:empirical_detection}.

\begin{table}[t]
\centering
\caption{Direct detection on C4 using Llama3.2-3B-Instruct. \ours{} improves the detection power of all three underlying watermarking schemes, especially under stringent FPR constraints.}
\vspace{-1.5mm}
\label{tab:main-c4-llama}
\setlength{\tabcolsep}{3.9pt}
\begin{tabular}{@{}l|cccc|cccc@{}}
\toprule
 & \multicolumn{4}{c|}{200 Tokens} & \multicolumn{4}{c}{400 Tokens} \\ \midrule
 & \multicolumn{3}{c|}{TPR@FPR=} & \multirow{2}{*}{\begin{tabular}[c]{@{}c@{}}Median \\ $p$-value $\downarrow$\end{tabular}} & \multicolumn{3}{c|}{TPR@FPR=} & \multirow{2}{*}{\begin{tabular}[c]{@{}c@{}}Median \\ $p$-value $\downarrow$\end{tabular}} \\ \cmidrule(lr){2-4} \cmidrule(lr){6-8}
 & 0.1\% $\uparrow$ & 0.01\% $\uparrow$ & \multicolumn{1}{c|}{0.001\% $\uparrow$} &  & 0.1\% $\uparrow$ & 0.01\% $\uparrow$ & \multicolumn{1}{c|}{0.001\% $\uparrow$} &  \\ \midrule
ENS-DiPmark & 47.0\% & 32.8\% & \multicolumn{1}{c|}{23.9\%} & 1.89e-3 & 73.8\% & 61.2\% & \multicolumn{1}{c|}{46.8\%} & 1.63e-5 \\
+ \ours & \textbf{68.3\%} & \textbf{58.9\%} & \multicolumn{1}{c|}{\textbf{47.2\%}} & \textbf{2.01e-5} & \textbf{94.5\%} & \textbf{88.6\%} & \multicolumn{1}{c|}{\textbf{81.7\%}} & \textbf{4.03e-10} \\ \midrule
SynthID-Text & 71.8\% & 60.5\% & \multicolumn{1}{c|}{50.3\%} & 9.59e-6 & 92.9\% & 86.5\% & \multicolumn{1}{c|}{80.1\%} & 1.73e-9 \\
+ \ours & \textbf{87.5\%} & \textbf{79.5\%} & \multicolumn{1}{c|}{\textbf{73.2\%}} & \textbf{5.79e-10} & \textbf{99.3\%} & \textbf{98.5\%} & \multicolumn{1}{c|}{\textbf{97.7\%}} & \textbf{8.25e-18} \\ \midrule
ENS-MCMark & 73.4\% & 62.5\% & \multicolumn{1}{c|}{54.0\%} & 3.86e-6 & 95.4\% & 91.3\% & \multicolumn{1}{c|}{83.5\%} & 2.00e-10 \\
+ \ours & \textbf{92.0\%} & \textbf{85.7\%} & \multicolumn{1}{c|}{\textbf{81.8\%}} & \textbf{1.84e-11} & \textbf{99.6\%} & \textbf{99.0\%} & \multicolumn{1}{c|}{\textbf{98.5\%}} & \textbf{9.36e-21} \\ \bottomrule
\end{tabular}
\end{table}

\vspace{-1mm}
\section{Experiments}
\vspace{-1mm}

\subsection{Experimental Setup}
\vspace{-1mm}

Our experiments focus on three representative distortion-free watermarks: ENS-DiPmark~\citep{wu2023dipmark}, SynthID-Text~\citep{dathathri2024scalable}, and ENS-MCMark~\citep{chen2026more}. For each baseline, we strictly follow the original watermark parameters and replace the single-token generation and scoring with \ours. Unless otherwise specified, generations use Top-\(p=0.95\), Top-\(k=50\), temperature \(1.0\), and a target length of 200 tokens. The skip penalty $\beta$ is set to 5.

We evaluate on instruction-tuned models from different families, including Llama3.2-3B-Instruct~\citep{touvron2023llama2}, Mistral-v0.3-7B-Instruct~\citep{jiang2023mistral}, and Qwen2.5-7B-Instruct~\citep{qwen2.5}.
For datasets, following prior works~\citep{hu2023unbiased,kirchenbauer2023watermark}, we use C4~\citep{raffel2020exploring}, MMW Book Report~\citep{piet2023mark}, Dolly Creative Writing~\citep{DatabricksBlog2023DollyV2}, and Longform QA~\citep{tu2023waterbench}.

We report true positive rates at fixed false positive rates (TPR@FPR), focusing on stringent false positive regimes that are most relevant for real-world deployment.
We additionally report the median detection \(p\)-value over watermarked samples.
Lower median \(p\)-values indicate stronger separation between watermarked and non-watermarked text.
\vspace{-1mm}
\subsection{Main Results}
\vspace{-1mm}
Table~\ref{tab:main-c4-llama} reports direct detection results on C4 using Llama3.2-3B-Instruct.
Across all three watermark families and both sequence lengths, \ours{} improves detection performance without changing the base watermark algorithm.
The gains are particularly pronounced at low false positive rates.
For 200-token generations, TPR@FPR \(=0.001\%\) increases from \(23.9\%\) to \(47.2\%\) for ENS-DiPmark, from \(50.3\%\) to \(73.2\%\) for SynthID-Text, and from \(54.0\%\) to \(81.8\%\) for ENS-MCMark.
At 400 tokens, the tandem variants reach \(81.7\%\), \(97.7\%\), and \(98.5\%\) TPR at the same false-positive rate, respectively.
The median \(p\)-values also decrease by several orders of magnitude, indicating that the improvement is not limited to a particular operating threshold.

\begin{figure}[t]
  \centering
  \begin{subfigure}[b]{0.32\textwidth}
    \centering
    \includegraphics[width=\textwidth]{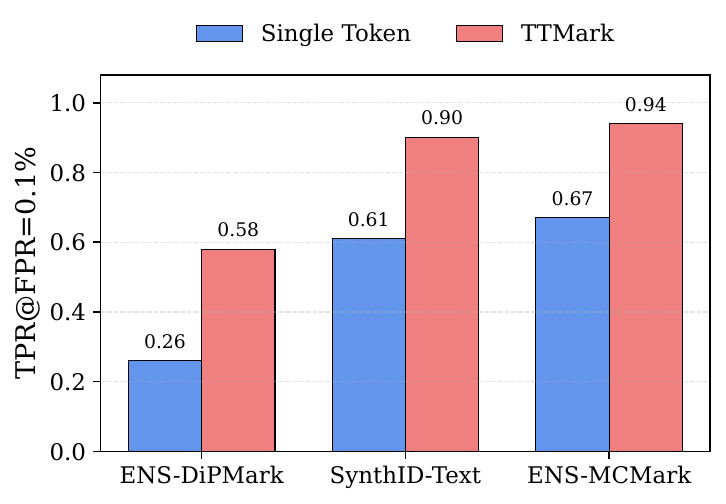}
    \caption{MMW Book Report}
    \label{fig:dataset-mmw-book}
  \end{subfigure}
  \hfill
  \begin{subfigure}[b]{0.32\textwidth}
    \centering
    \includegraphics[width=\textwidth]{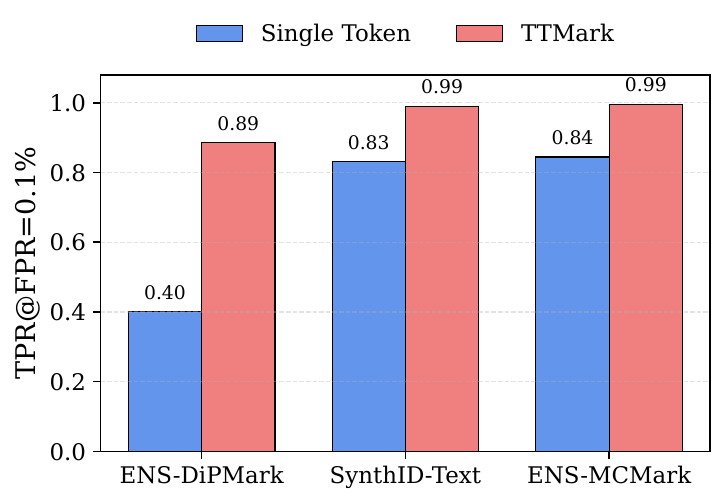}
    \caption{Longform QA}
    \label{fig:dataset-longform-qa}
  \end{subfigure}
  \hfill
  \begin{subfigure}[b]{0.32\textwidth}
    \centering
    \includegraphics[width=\textwidth]{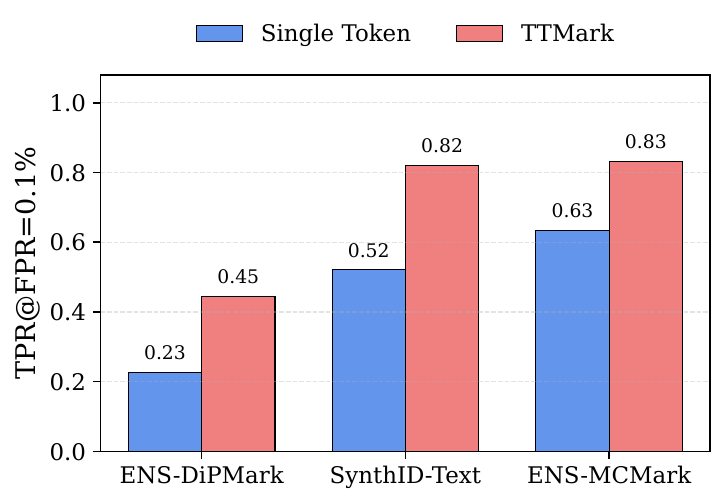}
    \caption{Dolly CW}
    \label{fig:dataset-dolly-cw}
  \end{subfigure}
  \vspace{-1mm}
  \caption{Cross-dataset direct detection performance on Llama3.2-3B-Instruct. Each panel reports TPR@FPR \(=0.1\%\) for single-token watermarking and its tandem counterpart.}
  \vspace{-3mm}
  \label{fig:dataset-comparison}
\end{figure}

Figure~\ref{fig:dataset-comparison} shows that the gains transfer beyond C4.
Across MMW Book Report, Long-form QA, and Dolly CW, \ours{} consistently improves TPR@FPR \(=0.1\%\) for all three watermark families, with especially clear gains on the more challenging MMW Book Report and Dolly Creative Writing tasks.
Additional results across datasets and model families are reported in Appendix~\ref{app:additional-experiments}.

\begin{figure}[t]
  \centering
  \begin{subfigure}[b]{0.32\textwidth}
    \centering
    \includegraphics[width=\textwidth]{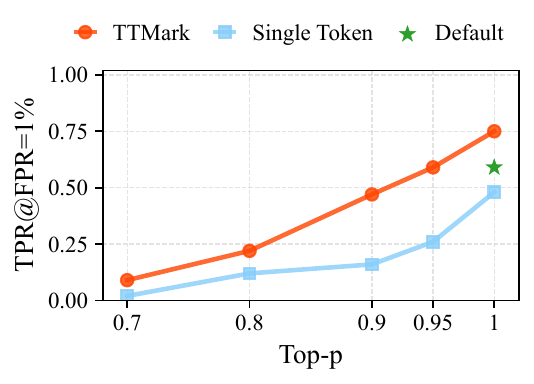}
    \caption{ENS-DiPmark}
    \label{fig:top-p-dipmark}
  \end{subfigure}
  \hfill
  \begin{subfigure}[b]{0.32\textwidth}
    \centering
    \includegraphics[width=\textwidth]{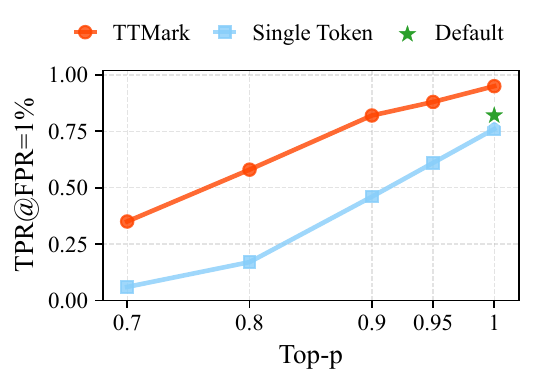}
    \caption{SynthID-Text}
    \label{fig:top-p-synthid}
  \end{subfigure}
  \hfill
  \begin{subfigure}[b]{0.32\textwidth}
    \centering
    \includegraphics[width=\textwidth]{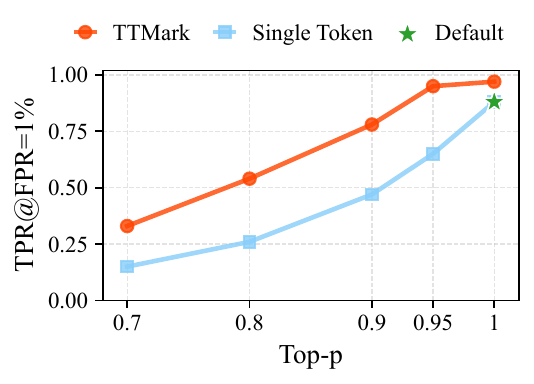}
    \caption{ENS-MCMark}
    \label{fig:top-p-mcmark}
  \end{subfigure}
  \vspace{-2mm}
  \caption{Effect of Top-\(p\) on detection performance for MMW Book Report using Llama3.2-3B-Instruct. Both tandem and single-token variants use Top-\(k=50\); ``Default'' denotes the setting without Top-\(k\) truncation.}
  \label{fig:top-p-ablation}
  \vspace{-2mm}
\end{figure}
\vspace{-1mm}
\section{Analysis}
\vspace{-1mm}
In this section, we analyze \ours{} under varying decoding entropy, stricter sampling parameters, computational costs, editing attacks, and localized detection settings. Additional experiments, including the skip-penalty \(\beta\) ablation, cross-model evaluations, generation-quality measurements, and extension to 3-token blocks, are provided in Appendix~\ref{app:additional-experiments}.
\vspace{-1mm}
\subsection{Ablation Study}\label{sec:analysis-ablation}

\vspace{-1mm}
\paragraph{Effect of Decoding Entropy.} We vary Top-\(p\) while keeping Top-\(k=50\) fixed on the MMW Book Report dataset with Llama3.2-3B-Instruct, to evaluate the performance under different decoding entropy.
Figure~\ref{fig:top-p-ablation} compares the tandem and single-token variants for each base watermark.
The tandem variants maintain stronger detection across the evaluated Top-\(p\) range, showing that the benefit of \ours{} persists under changes in decoding entropy.

\vspace{-1mm}
\paragraph{Stricter Sampling Parameters.}
Practical deployments often use more conservative decoding than the default experimental setting.
For example, code generation, tool-using agents, and retrieval-augmented generation commonly reduce temperature in order to improve determinism and factual consistency.
In the appendix, Table~\ref{tab:suggested-sampling-parameters} reports representative recommended sampling parameters from widely used model families.
Such settings are challenging for watermarking because the entropy becomes much smaller.
Table~\ref{tab:strict-decoding} evaluates this low-entropy regime using Qwen2.5-7B-Instruct on MMW Book Report with the recommended sampling parameters. \ours\ consistently improves TPR at all reported false-positive rates, with negligible runtime difference.
For SynthID-Text and ENS-MCMark, \ours{} more than doubles TPR@FPR \(=0.001\%\), from \(8\%\) to \(20\%\) and from \(5\%\) to \(28\%\), respectively.

\begin{table}[t]
\centering
\caption{Direct detection under stricter sampling parameters on MMW Book Report using Qwen2.5-7B-Instruct. Generations use the recommended sampling parameters with Top-\(p=0.8\), Top-\(k=20\), and temperature \(0.7\). Even when the effective candidate support is small, \ours{} improves detection with comparable generation time.}
\vspace{-1mm}
\label{tab:strict-decoding}
\begin{tabular}{@{}l|c|ccc|c|c@{}}
\toprule
 & \multirow{2}{*}{\begin{tabular}[c]{@{}c@{}}Average \\Token Counts\end{tabular}} & \multicolumn{3}{c|}{TPR@FPR=} & \multirow{2}{*}{\begin{tabular}[c]{@{}c@{}}Median \\ $p$-value $\downarrow$\end{tabular}} & \multirow{2}{*}{\begin{tabular}[c]{@{}c@{}}Seconds per\\ 1k Tokens $\downarrow$\end{tabular}} \\ \cmidrule(lr){3-5}
 &  & 0.1\% $\uparrow$ & 0.01\% $\uparrow$ & 0.001\% $\uparrow$ &  &  \\ \midrule
ENS-DiPmark & 1.37 & 9\% & 2\% & 0\% & 1.86e-1 & \textbf{0.79} \\
+ \ours & 1.93 & \textbf{18\%} & \textbf{9\%} & \textbf{2\%} & \textbf{3.46e-2} & 0.80 \\ \midrule
SynthID-Text & 1.39 & 21\% & 14\% & 8\% & 1.46e-2 & 0.86 \\
+ \ours & 1.93 & \textbf{52\%} & \textbf{36\%} & \textbf{20\%} & \textbf{7.96e-4} & \textbf{0.79} \\ \midrule
ENS-MCMark & 1.38 & 26\% & 13\% & 5\% & 1.27e-2 & 0.87 \\
+ \ours & 1.91 & \textbf{58\%} & \textbf{41\%} & \textbf{28\%} & \textbf{2.82e-4} & \textbf{0.80} \\ \bottomrule
\end{tabular}
\vspace{-3mm}
\end{table}

\begin{figure}[t]
  \centering
  \begin{subfigure}[b]{0.32\textwidth}
    \centering
    \includegraphics[width=\textwidth]{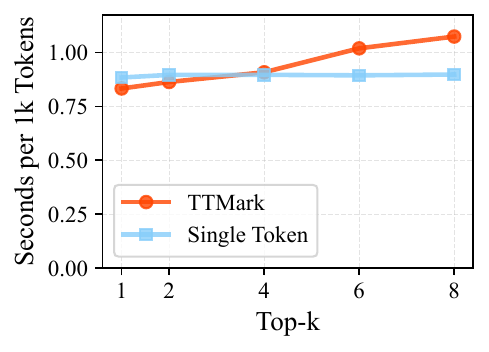}
    \caption{ENS-DiPmark}
    \label{fig:runtime-dipmark}
  \end{subfigure}
  \hfill
  \begin{subfigure}[b]{0.32\textwidth}
    \centering
    \includegraphics[width=\textwidth]{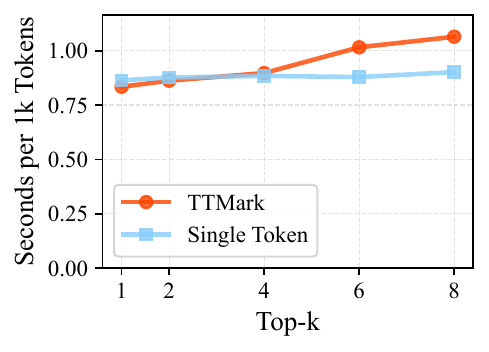}
    \caption{SynthID-Text}
    \label{fig:runtime-synthid}
  \end{subfigure}
  \hfill
  \begin{subfigure}[b]{0.32\textwidth}
    \centering
    \includegraphics[width=\textwidth]{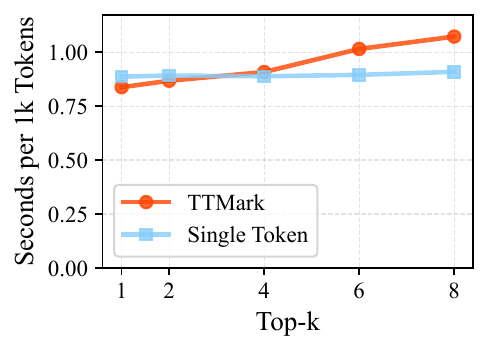}
    \caption{ENS-MCMark}
    \label{fig:runtime-mcmark}
  \end{subfigure}
  \vspace{-1mm}
  \caption{Runtime of tandem generation as a function of the effective candidate support on MMW Book Report using Llama3.2-3B-Instruct. Measurements are taken with batch size 32. Branch-isolating generation keeps the overhead modest because all first-token branches are evaluated together, and memory bandwidth is the main bottleneck. The accelerated generation speed when $m$ is small stems from a 50\% reduction in watermarking function calls.}
  \vspace{-3mm}
  \label{fig:runtime-support}
\end{figure}

\vspace{-2mm}
\paragraph{Computational Cost.} This overhead of \ours\ is bounded by the size of the effective support $m$ after decoding.
In practice, truncation makes this support small, and the branch-isolating generation procedure in Sec.~\ref{sec:generation} evaluates all candidate branches in a single batched forward pass.
Figure~\ref{fig:runtime-support} reports the resulting runtime as a function of the number of remaining candidates $m$.
The runtime increases smoothly with support size, but remains close to the single-token baselines in the operating regimes used in our experiments. Because memory bandwidth is the primary bottleneck, the observed overhead is much smaller than the $(m+1)/2$ ratio discussed in Sec.~\ref{sec:generation}. Furthermore, the accelerated generation speed when $m$ is small stems from a 50\% reduction in watermarking function calls.

\vspace{-2mm}
\subsection{Generation Quality}\label{sec:analysis-quality}
\vspace{-2mm}
Because \ours{} is distortion-free at the token-pair level, it can generally preserve the output quality.
We empirically verify this expectation in Table~\ref{tab:generation-quality} on summarization and machine translation benchmarks.
Across ROUGE, BLEU, and BERTScore, the tandem variants remain generally the same compared with the corresponding single-token watermarks and the unwatermarked model.

\vspace{-2mm}
\subsection{Robustness to Text Editing}\label{sec:analysis-robustness}
\vspace{-2mm}

We next evaluate whether the stronger detection signal persists after edits.
Table~\ref{tab:robustness-attacks} reports detection after rephrasing and back-translation on C4 using GPT5-Mini.
For these experiments, \(30\%\) of each generated passage is edited, and detection uses the empirical alignment detection procedure from Sec.~\ref{sec:detection}.
Across both attack types, \ours{} improves TPR at fixed FPR and AUROC for all three underlying watermark families.
Under rephrasing, \ours{} improves TPR at FPR$=0.1\%$ by \(16.1\), \(13.3\), and \(18.8\) percentage points for ENS-DiPmark, SynthID-Text, and ENS-MCMark, respectively.
Under back translation, the corresponding gains are \(13.8\), \(12.3\), and \(16.4\) percentage points.
These results suggest that the additional pairwise signal remains useful even when some token boundaries are corrupted, provided the detector is allowed to search over plausible local alignments.

\begin{table}[t]
\centering
\caption{Robustness to rephrasing and back-translation on C4 using GPT5-Mini. The generation model is Llama3.2-3B-Instruct. Each attack edits \(30\%\) of the generated passage. \ours{} improves detection after edits across all base watermarking schemes.}
\vspace{-1.5mm}
\label{tab:robustness-attacks}
\setlength{\tabcolsep}{5pt}
\begin{tabular}{@{}l|ccc|ccc@{}}
\toprule
 & \multicolumn{3}{c|}{Rephrase} & \multicolumn{3}{c}{Back Translation} \\ \cmidrule(l){2-7} 
 & FPR=1\%$\uparrow$ & \multicolumn{1}{c|}{FPR=0.1\% $\uparrow$} & AUROC $\uparrow$ & FPR=1\%$\uparrow$ & \multicolumn{1}{c|}{FPR=0.1\% $\uparrow$} & AUROC $\uparrow$ \\ \midrule
ENS-DiPmark & 59.4\% & \multicolumn{1}{c|}{34.6\%} & 0.9320 & 55.0\% & \multicolumn{1}{c|}{38.0\%} & 0.9486 \\
+ \ours & \textbf{69.8\%} & \multicolumn{1}{c|}{\textbf{50.7\%}} & \textbf{0.9449} & \textbf{72.0\%} & \multicolumn{1}{c|}{\textbf{51.8\%}} & \textbf{0.9532} \\ \midrule
SynthID-Text & 72.2\% & \multicolumn{1}{c|}{59.0\%} & 0.9529 & 75.6\% & \multicolumn{1}{c|}{60.0\%} & 0.9646 \\
+ \ours & \textbf{80.0\%} & \multicolumn{1}{c|}{\textbf{72.3\%}} & \textbf{0.9570} & \textbf{83.1\%} & \multicolumn{1}{c|}{\textbf{72.3\%}} & \textbf{0.9726} \\ \midrule
ENS-MCMark & 75.5\% & \multicolumn{1}{c|}{61.9\%} & 0.9587 & 78.0\% & \multicolumn{1}{c|}{64.9\%} & 0.9687 \\
+ \ours & \textbf{88.0\%} & \multicolumn{1}{c|}{\textbf{80.7\%}} & \textbf{0.9759} & \textbf{89.3\%} & \multicolumn{1}{c|}{\textbf{81.3\%}} & \textbf{0.9795} \\ \bottomrule
\end{tabular}
\end{table}

\vspace{-2mm}
\subsection{Localized Watermark Detection}\label{sec:analysis-localized}
\vspace{-2mm}
Finally, we consider a localized detection setting in which only a contiguous span of a document is watermarked.
The detector must both decide whether a watermark is present and identify the watermarked region.
For this evaluation, we strictly follow the settings of WaterSeeker~\citep{pan2025waterseeker}, including its datasets and hyperparameters.
Table~\ref{tab:localized-detection} reports the resulting localized-detection performance for Llama3.2-3B-Instruct and Qwen2.5-7B-Instruct.
Compared with the single-token baselines, \ours{} substantially reduces false negative rates while maintaining low FPR, leading to higher F1 and IoU. The consistent IoU improvements indicate that the stronger pairwise signal helps not only global detection but also finer-grained localization of watermarked spans.

\begin{table}[t]
\centering
\caption{Localized watermark detection. The protocol evaluates whether the detector can recover the watermarked span within a longer document. \ours{} reduces false negatives and improves F1 and IoU on both model families.}
\vspace{-1.5mm}
\label{tab:localized-detection}
\begin{tabular}{@{}l|cccc|cccc@{}}
\toprule
 & \multicolumn{4}{c|}{Llama3.2-3B-Instruct} & \multicolumn{4}{c}{Qwen2.5-7B-Instruct} \\ \cmidrule(l){2-9} 
 & FPR$\downarrow$ & FNR$\downarrow$ & F1$\uparrow$ & IoU$\uparrow$ & FPR$\downarrow$ & FNR$\downarrow$ & F1$\uparrow$ & IoU$\uparrow$ \\ \midrule
ENS-DiPmark & \textbf{0.000} & 0.543 & 0.627 & 0.351 & 0.010 & 0.567 & 0.600 & 0.330 \\
+ \ours & \textbf{0.000} & \textbf{0.277} & \textbf{0.839} & \textbf{0.590} & \textbf{0.003} & \textbf{0.303} & \textbf{0.820} & \textbf{0.574} \\ \midrule
SynthID-Text & 0.017 & 0.230 & 0.862 & 0.618 & \textbf{0.007} & 0.213 & 0.877 & 0.623 \\
+ \ours & \textbf{0.010} & \textbf{0.050} & \textbf{0.969} & \textbf{0.785} & 0.013 & \textbf{0.063} & \textbf{0.961} & \textbf{0.787} \\ \midrule
ENS-MCMark & 0.006 & 0.210 & 0.879 & 0.618 & 0.037 & 0.153 & 0.899 & 0.637 \\
+ \ours & \textbf{0.003} & \textbf{0.070} & \textbf{0.962} & \textbf{0.750} & \textbf{0.013} & \textbf{0.043} & \textbf{0.973} & \textbf{0.791} \\ \bottomrule
\end{tabular}
\vspace{-2mm}
\end{table}

\vspace{-2mm}
\section{Conclusion}
\vspace{-2mm}
We introduce \ours, a novel framework that extends distortion-free watermarking from individual tokens to adjacent token pairs. By watermarking the joint distribution of consecutive tokens, the proposed method expands the effective watermarking alphabet to $V^{2}$, successfully exploiting both immediate token entropy and conditional entropy to overcome the inherent detection bottlenecks of single-token approaches. To operationalize this efficiently, we developed a branch-isolating concatenated tandem generation algorithm that constructs the required joint distribution in a single forward pass. Theoretical analyses formally establish that this pairwise approach yields superior expected detection strength, particularly within challenging low-entropy regimes. Ultimately, extensive empirical evaluations across multiple language models and datasets confirm that \ours{} consistently enhances general detectability, improves robustness against semantic-preserving edits, and substantially boosts localized detection capabilities, all while preserving the underlying generation quality.

\clearpage

\subsection*{AI use statement}


We used generative AI tools to assist with coding and debugging experimental implementations, as well as to polish and refine the writing of the manuscript. Generative AI tools were not used to generate, develop, or formalize the research ideas or scientific contributions of this work. All AI-assisted code and text were reviewed and verified by the authors. The authors take full responsibility for the research methodology, experimental results, and final content of this paper.








\bibliography{iclr2027_conference}
\bibliographystyle{iclr2027_conference}

\appendix

\section{Missing Proofs}\label{app:proofs}

\subsection{Proof of Theorem~\ref{thm:single-step-z-bound}}\label{app:proof-single-step-z-bound}

\begin{proof}
Recall that the normalized score is defined as \(z(x,k)=(s(x,k)-\mu)/\sigma\).
For the model-agnostic scoring rules considered in the main text, the null score has the token-independent mean
\(\mathbb{E}_{k}\left[s(x,k)\right]=\mu\) and variance \(\operatorname{Var}_{k}[s(x,k)]=\sigma^2\) for every \(x\).
Therefore, for each fixed token \(x\),
\begin{equation}
    \mathbb{E}_{k}\left[z(x,k)\right]
    =
    \frac{\mathbb{E}_{k}\left[s(x,k)\right]-\mu}{\sigma}
    =
    0,
    \qquad
    \operatorname{Var}_{k}[z(x,k)] = 1.
\end{equation}
Using this zero-mean property together with distortion-freeness,
\(\mathbb{E}_{k}\left[P_w(x\mid k)\right]=P(x)\), we can center the watermarked probability inside the expectation:
\begin{align}
    \mathbb{E}_{k,\,x\sim P_w(\cdot\mid k)}\left[z(x,k)\right]
    &=
    \sum_{x\in V}
    \mathbb{E}_{k}\left[P_w(x\mid k)z(x,k)\right] \nonumber\\
    &=
    \sum_{x\in V}
    \left(
    \mathbb{E}_{k}\left[(P_w(x\mid k)-P(x))z(x,k)\right]
    +
    P(x)\mathbb{E}_{k}\left[z(x,k)\right]
    \right) \nonumber\\
    &=
    \sum_{x\in V}
    \mathbb{E}_{k}\left[(P_w(x\mid k)-P(x))z(x,k)\right].
\end{align}
For each \(x\), define the centered random variables
\(A_x(k)=P_w(x\mid k)-P(x)\) and \(B_x(k)=z(x,k)\).
Then \(\mathbb{E}_{k}\left[A_x(k)\right]=0\), \(\mathbb{E}_{k}\left[B_x(k)\right]=0\), and
\(\mathbb{E}_{k}\left[B_x(k)^2\right]=\operatorname{Var}_{k}[z(x,k)]=1\).
Applying Cauchy--Schwarz to these two random variables gives
\begin{align}
    \mathbb{E}_{k}\left[(P_w(x\mid k)-P(x))z(x,k)\right]
    &\leq
    \left|
    \mathbb{E}_{k}\left[A_x(k)B_x(k)\right]
    \right| \nonumber\\
    &\leq
    \sqrt{\mathbb{E}_{k}\left[A_x(k)^2\right]}
    \sqrt{\mathbb{E}_{k}\left[B_x(k)^2\right]} \nonumber\\
    &=
    \sqrt{\operatorname{Var}_{k}[P_w(x\mid k)]}.
\end{align}
It remains to bound the first variance term.
Since \(0\leq P_w(x\mid k)\leq 1\), we have \(P_w(x\mid k)^2\leq P_w(x\mid k)\).
Together with \(\mathbb{E}_{k}\left[P_w(x\mid k)\right]=P(x)\), this implies
\begin{equation}
    \operatorname{Var}_{k}[P_w(x\mid k)]
    =
    \mathbb{E}_{k}\left[P_w(x\mid k)^2\right] - P(x)^2
    \leq
    P(x)-P(x)^2.
\end{equation}
Substituting this bound into the previous inequality and summing over \(x\in V\) yields
\[
    \mathbb{E}_{k\sim P_{\mathcal{K}},\,x\sim P_w(\cdot\mid k)}\left[z(x,k)\right]
    \leq
    \sum_{x\in V}\sqrt{P(x)-P(x)^2},
\]
which proves the claim.
\end{proof}

\subsection{Proof of Theorem~\ref{thm:pairwise-z-bound}}\label{app:proof-pairwise-z-bound}

\begin{proof}
We prove the claim by comparing the contribution of each fixed first token \(x_1\).
Fix \(x_1\), write \(a=P_1(x_1)\), and let \(b_y=P_{2\mid 1=x_1}(y)\) for \(y\in V\).
If \(a=0\), then this \(x_1\) contributes zero to all three quantities in the desired inequality, so the claim is trivial.
We therefore assume \(a>0\), in which case \(b=(b_y)_{y\in V}\) is a probability distribution.

The contribution of this \(x_1\) to \(f(P_{12})\) is
\[
    S
    =
    \sum_{y\in V}\sqrt{P_{12}(x_1,y)-P_{12}(x_1,y)^2}
    =
    \sum_{y\in V}\sqrt{a b_y(1-a b_y)}.
\]
The corresponding contribution to \(f(P_1)\) is
\[
    A=\sqrt{P_1(x_1)-P_1(x_1)^2}=\sqrt{a(1-a)},
\]
and the contribution to \(\mathbb{E}_{x_1\sim P_1}f(P_{2\mid 1=x_1})\) is
\[
    B=a\sum_{y\in V}\sqrt{b_y(1-b_y)}.
\]
It suffices to show the pointwise inequality
\begin{equation}\label{eq:pointwise-pair-bound}
    S\ge \frac{A+B}{\sqrt{2}}.
\end{equation}

We start by expanding \(S^2\). Its diagonal terms are
\[
    \sum_y a b_y(1-a b_y)
    =
    a - a^2\sum_y b_y^2
    =
    a(1-a)+a^2\sum_y b_y(1-b_y).
\]
Here we used \(\sum_y b_y=1\).
The first term is \(A^2\), while the second term is the diagonal part of \(B^2\).
It remains to compare the cross terms.
The cross terms of \(S^2\) are
\[
    2a\sum_{y<y'}
    \sqrt{
        b_y b_{y'}(1-a b_y)(1-a b_{y'})
    },
\]
whereas the cross terms of \(B^2\) are
\[
    2a^2\sum_{y<y'}
    \sqrt{
        b_y b_{y'}(1-b_y)(1-b_{y'})
    }.
\]
For any pair \(y\neq y'\), because \(a\in[0,1]\) and \(b_y,b_{y'}\in[0,1]\),
\[
    (1-a b_y)(1-a b_{y'})
    -
    a^2(1-b_y)(1-b_{y'})
    =
    (1-a)\left(1+a-a(b_y+b_{y'})\right)
    \geq 0.
\]
Thus
\[
    a\sqrt{
        b_y b_{y'}(1-a b_y)(1-a b_{y'})
    }
    \ge
    a^2\sqrt{
        b_y b_{y'}(1-b_y)(1-b_{y'})
    },
\]
so every cross term of \(S^2\) is at least the corresponding cross term of \(B^2\).
Combining the diagonal and cross-term comparisons gives
\[
    S^2 \ge A^2+B^2.
\]
Finally, since \(A,B\ge0\), we have
\[
    A^2+B^2 \ge \frac{(A+B)^2}{2},
\]
which proves Eq.~\eqref{eq:pointwise-pair-bound}.
Summing Eq.~\eqref{eq:pointwise-pair-bound} over all \(x_1\in V\) yields
\[
    \sum_{x_1\in V}S
    \ge
    \frac{
    \sum_{x_1\in V}A+
    \sum_{x_1\in V}B
    }{\sqrt{2}}.
\]
By construction,
\(\sum_{x_1\in V}S=f(P_{12})\),
\(\sum_{x_1\in V}A=f(P_1)\), and
\(\sum_{x_1\in V}B=\mathbb{E}_{x_1\sim P_1}f(P_{2\mid 1=x_1})\).
Substituting these identities gives the desired inequality.
\end{proof}

\section{Algorithms for Generation and Detection}

Algorithm~\ref{alg:generation} constructs the filtered two-token distribution by evaluating all admissible first-token branches in a single masked forward pass.
The branch-isolating attention mask ensures that each candidate continuation is equivalent to an independent cached decoding step from the same prefix, while allowing the implementation to exploit batching on modern accelerators.
Algorithm~\ref{alg:empirical_detection} gives the dynamic program used for edited or partially shifted text.

\begin{algorithm}[htbp]
  \caption{Branch-Isolating Concatenated Tandem Generation}
  \label{alg:generation}
  \begin{algorithmic}[1]
    \Require Language model \(M\); prompt \(p\); generation length \(T\); decoding parameters \(\mathrm{Params}\); pairwise watermarking rule defining \(P_\mathrm{TTMark}\); watermark keys \(\{k_t\}_{t\ge 0}\)
    \State \(\ell,\mathrm{KV}\gets M(p)\) \Comment{Logits for \(P_M(\cdot\mid p)\) and prefix cache}
    \State \(t\gets 0\), \(\bm{x}_{1:0}\gets [\,]\)
    \While{\(t<T\)}
      \State \(P_1 \gets \operatorname{Filter}(\operatorname{softmax}(\ell),\mathrm{Params})\)
      \If{\(t=T-1\)}
        \State Sample \(x_{t+1}\sim P_1\) and set \(\bm{x}_{1:t+1}\gets[\bm{x}_{1:t},x_{t+1}]\)
        \State \textbf{break}
      \EndIf
      \State \(\mathcal{S}_1=\{x_{t+1}^{(i)}\}_{i=1}^{m}\gets\{x\in V:P_1(x)>0\}\) \Comment{Candidate support after truncation}
      \State \(\bm{x}_{t+1}^{(1:m)}\gets[x_{t+1}^{(1)},\ldots,x_{t+1}^{(m)}]\)
      \State \(L\gets\operatorname{length}(\mathrm{KV})\) \Comment{Current prefix-cache length}
      \State Construct the branch-isolating attention mask \(A\) by Eq.~\ref{eq:masked_generation} and assign position index \(L+1\) to all candidates
      \State \(\{\ell_2^{(i)},\mathrm{KV}^{(i)}\}_{i=1}^{m}\gets M(\bm{x}_{t+1}^{(1:m)}\mid \mathrm{KV},A)\)
      \For{\(i=1,\ldots,m\)}
        \State \(P_2^{(i)}\gets \operatorname{Filter}(\operatorname{softmax}(\ell_2^{(i)}),\mathrm{Params})\) \Comment{Apply the same decoding parameters to \(x_{t+2}\)}
        \State \(\mathcal{S}_2^{(i)}\gets\{x\in V:P_2^{(i)}(x)>0\}\)
      \EndFor
      \State \(\mathcal{S}_{12}\gets\{(x_{t+1}^{(i)},x):x_{t+1}^{(i)}\in\mathcal{S}_1,\;x\in\mathcal{S}_2^{(i)}\}\)
      \State Define the pre-watermark pair distribution on \(\mathcal{S}_{12}\) using \(P_1(x_{t+1}^{(i)})P_2^{(i)}(x_{t+2})\), following Eq.~\ref{eq:joint-next-token-factorization}
      \State Obtain \(P_\mathrm{TTMark}(x_{t+1},x_{t+2}\mid \bm{x}_{1:t},p,k_t)\) by reweighting the pre-watermark pair distribution
      \State Sample \((x_{t+1},x_{t+2})\sim P_\mathrm{TTMark}(x_{t+1},x_{t+2}\mid \bm{x}_{1:t},p,k_t)\)
      \State \(\bm{x}_{1:t+2}\gets[\bm{x}_{1:t},x_{t+1},x_{t+2}]\)
      \State \(\mathrm{KV}^{\star}\gets \mathrm{KV}^{(i^\star)}\), where \(x_{t+1}=x_{t+1}^{(i^\star)}\)
      \State \(\ell,\mathrm{KV}\gets M(x_{t+2}\mid \mathrm{KV}^{\star})\)
      \State \(t\gets t+2\)
    \EndWhile
    \State \Return Generated sequence \(\bm{x}_{1:T}\)
  \end{algorithmic}
\end{algorithm}

\begin{algorithm}[htbp]
  \caption{Empirical Detection}
  \label{alg:empirical_detection}
  \begin{algorithmic}[1]
    \Require Observed tokens \(\bm{y}_{1:L}\); pairwise score \(s\); keys \(\{k_i\}\); null mean \(\mu\); null standard deviation \(\sigma\); skip penalty \(\beta\)
    \Ensure Empirical detection statistic \(z_{\mathrm{emp}}\)
    \State \(n\gets\floor{L/2}\)
    \State Initialize \(D_t\gets-\infty\) for \(0\le t\le L\)
    \State \(D_0\gets 0\)
    \For{\(t=1,\ldots,L\)}
      \State \(D_t\gets D_{t-1}-\beta\) \Comment{Skip \(y_t\)}
      \If{\(t\ge 2\)}
        \State \(r\gets (s((y_{t-1},y_t),k_{t-1})-\mu)/\sigma\)
        \State \(D_t\gets \max(D_t, D_{t-2}+r)\) \Comment{Retain adjacent pair}
      \EndIf
    \EndFor
    \State \Return \(z_{\mathrm{emp}}\gets D_L/\sqrt{n}\)
  \end{algorithmic}
\end{algorithm}

\section{Additional Experiments}\label{app:additional-experiments}

\subsection{Ablation of Skip Penalty \texorpdfstring{\(\beta\)}{beta}}

\begin{figure}[t]
  \centering
  \begin{subfigure}[b]{0.49\textwidth}
    \centering
    \includegraphics[width=\textwidth]{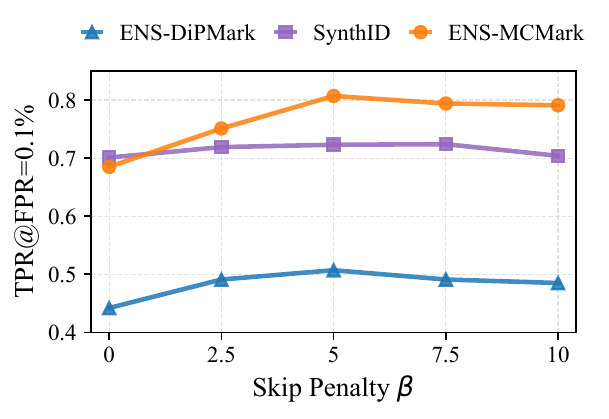}
    \caption{Rephrasing}
    \label{fig:beta-rephrase}
  \end{subfigure}
  \hfill
  \begin{subfigure}[b]{0.49\textwidth}
    \centering
    \includegraphics[width=\textwidth]{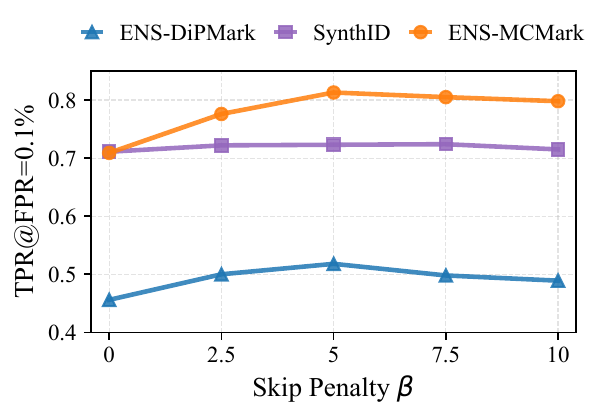}
    \caption{Back translation}
    \label{fig:beta-back-translation}
  \end{subfigure}
  \caption{Ablation of the empirical detector skip penalty \(\beta\) under edits. Moderate penalties balance alignment flexibility against over-selection of isolated high-scoring pairs.}
  \label{fig:ablation_beta}
\end{figure}

The skip penalty \(\beta\) controls the trade-off in the empirical alignment detector.
When \(\beta\) is too small, the detector can skip many tokens and select a small number of unusually high-scoring adjacent pairs, which weakens separation after null calibration.
When \(\beta\) is too large, the detector becomes close to direct detection and loses the ability to recover from local insertions, deletions, or modifications.
Figure~\ref{fig:ablation_beta} explores the effect of \(\beta\) under rephrasing and back-translation attacks.
The results show that moderate skip penalties provide the most reliable behavior across both perturbation types.
We therefore use \(\beta=5\) in the main experiments, which preserves most of the pairwise watermark signal while still allowing the detector to realign around locally edited spans.

\subsection{Cross-Dataset Results on Additional Model Families}

\begin{figure}[htbp]
  \centering
  \begin{subfigure}[b]{0.32\textwidth}
    \centering
    \includegraphics[width=\textwidth]{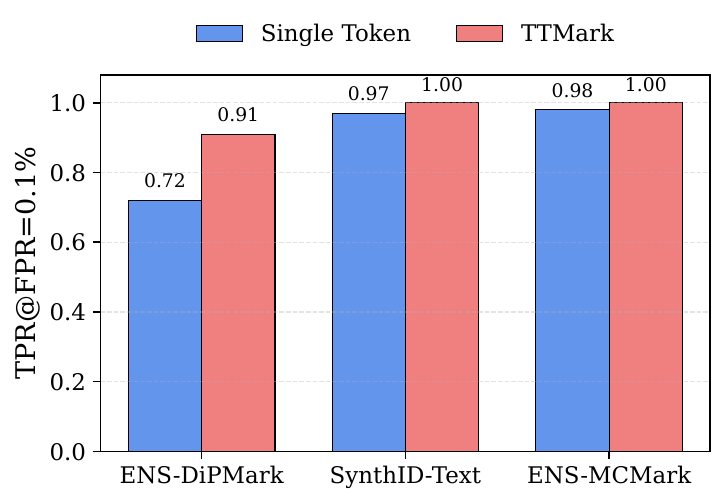}
    \caption{MMW Book Report}
    \label{fig:qwen-mmw-book}
  \end{subfigure}
  \hfill
  \begin{subfigure}[b]{0.32\textwidth}
    \centering
    \includegraphics[width=\textwidth]{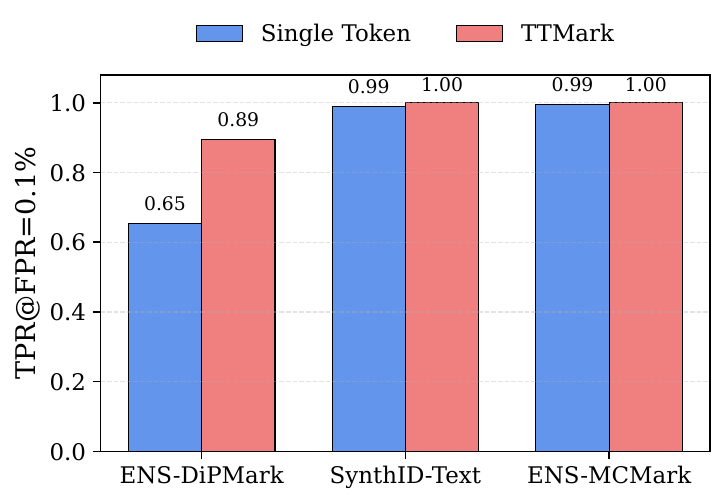}
    \caption{Longform QA}
    \label{fig:qwen-longform-qa}
  \end{subfigure}
  \hfill
  \begin{subfigure}[b]{0.32\textwidth}
    \centering
    \includegraphics[width=\textwidth]{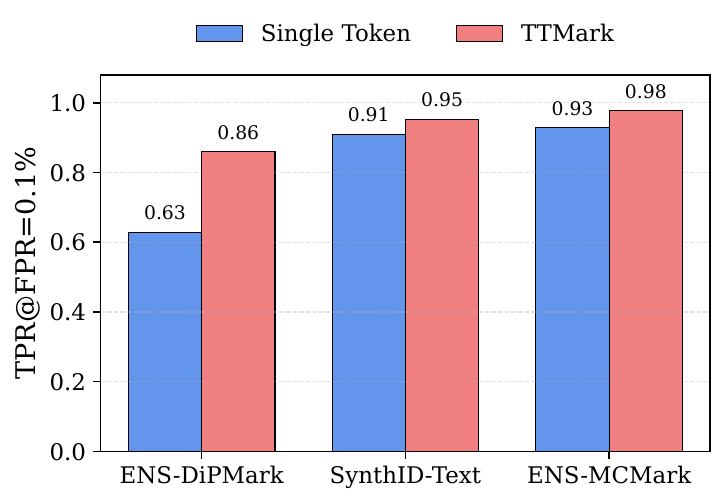}
    \caption{Dolly CW}
    \label{fig:qwen-dolly-cw}
  \end{subfigure}
  \caption{Cross-dataset direct detection performance on Qwen2.5-7B-Instruct. Each panel reports TPR@FPR \(=0.1\%\) for single-token watermarking and its tandem counterpart.}
  \label{fig:qwen-dataset-comparison}
\end{figure}

\begin{figure}[htbp]
  \centering
  \begin{subfigure}[b]{0.32\textwidth}
    \centering
    \includegraphics[width=\textwidth]{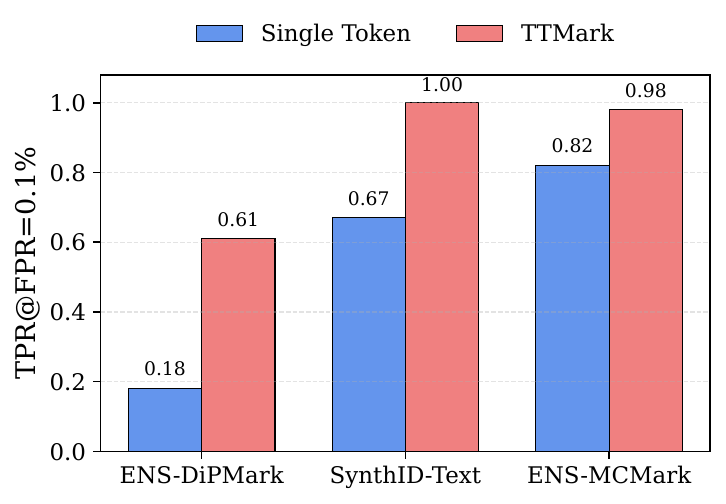}
    \caption{MMW Book Report}
    \label{fig:mistral-mmw-book}
  \end{subfigure}
  \hfill
  \begin{subfigure}[b]{0.32\textwidth}
    \centering
    \includegraphics[width=\textwidth]{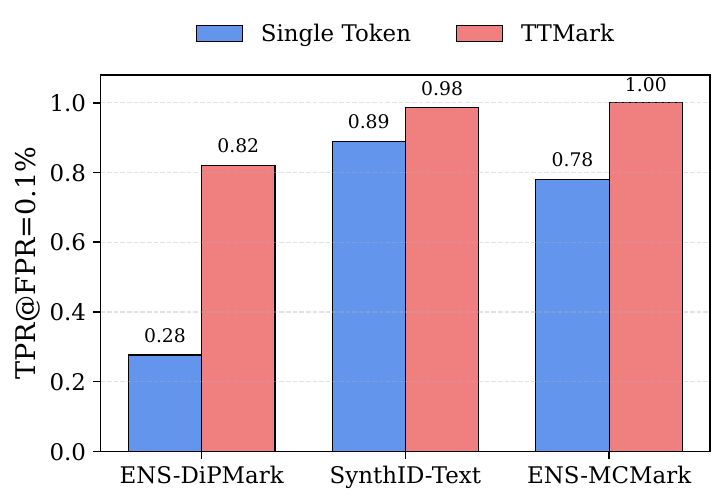}
    \caption{Longform QA}
    \label{fig:mistral-longform-qa}
  \end{subfigure}
  \hfill
  \begin{subfigure}[b]{0.32\textwidth}
    \centering
    \includegraphics[width=\textwidth]{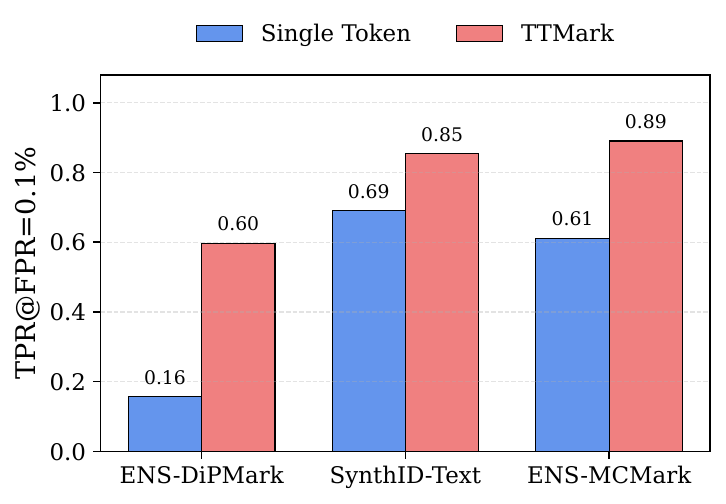}
    \caption{Dolly CW}
    \label{fig:mistral-dolly-cw}
  \end{subfigure}
  \caption{Cross-dataset direct detection performance on Mistral-7B-Instruct-v0.3. Each panel reports TPR@FPR \(=0.1\%\) for single-token watermarking and its tandem counterpart.}
  \label{fig:mistral-dataset-comparison}
\end{figure}

The main text reports cross-dataset results for Llama3.2-3B-Instruct.
Figures~\ref{fig:qwen-dataset-comparison} and~\ref{fig:mistral-dataset-comparison} provide the corresponding results for Qwen2.5-7B-Instruct and Mistral-7B-Instruct-v0.3.
Across MMW Book Report, Long-form QA, and Dolly Creative Writing, the tandem variants consistently improve direct detection at FPR \(=0.1\%\) over their single-token counterparts.
This pattern holds for all three underlying watermarking schemes, indicating that the gain from \ours\ is not specific to a single model family or task distribution.

\subsection{Sampling Parameters for Low-Entropy Decoding}

Table~\ref{tab:suggested-sampling-parameters} records representative recommended decoding parameters from common model families.
These settings are more conservative than the default experimental configuration and reduce the effective support size available to the watermark.
The dash in the Top-\(k\) column indicates that no explicit Top-\(k\) truncation is recommended in that setting.

\begin{table}[t]
\centering
\caption{Representative stricter sampling parameters recommended by model families. These settings reduce decoding entropy and can substantially decrease the effective candidate support available for watermarking. The dash in the Top-\(k\) column indicates that no explicit Top-\(k\) truncation is recommended in that setting.}
\label{tab:suggested-sampling-parameters}
\begin{tabular}{@{}l|ccc@{}}
\toprule
 & Top-P & Top-K & Temperature \\ \midrule
Qwen3 Series & 0.95 & 20 & 0.6 \\
Qwen2.5 Series & 0.8 & 20 & 0.7 \\
Llama3 Series & 0.9 & - & 0.6 \\ \bottomrule
\end{tabular}
\end{table}

\subsection{Generation Quality}

Because \ours{} is distortion-free over token pairs, it should not systematically change the marginal text distribution when averaged over the watermark key.
Table~\ref{tab:generation-quality} evaluates this expectation on summarization and machine translation.
Across ROUGE, BLEU, and BERTScore, the tandem variants remain close to both the unwatermarked model and the corresponding single-token watermarking schemes.

\begin{table}[t]
\centering
\caption{Generation quality under watermarking. The tandem variants preserve summarization and translation quality relative to the corresponding single-token watermarking schemes.}
\label{tab:generation-quality}
\begin{tabular}{@{}l|cccc|cc@{}}
\toprule
 & \multicolumn{4}{c|}{Text Summarization} & \multicolumn{2}{c}{Machine Translation} \\ \cmidrule(l){2-7} 
 & ROUGE-1 & ROUGE-2 & ROUGE-L & BERTScore & BLEU & BERTScore \\ \midrule
No Watermark & 0.3849 & 0.1465 & 0.2548 & 0.2733 & 21.30 & 0.8336 \\
+ \ours & 0.3854 & 0.1473 & 0.2554 & 0.2738 & 21.69 & 0.8388 \\ \midrule
ENS-DiPmark & 0.3855 & 0.1333 & 0.2432 & 0.2734 & 20.93 & 0.8332 \\
+ \ours & 0.3846 & 0.1333 & 0.2430 & 0.2726 & 21.55 & 0.8375 \\
SynthID-Text & 0.3850 & 0.1471 & 0.2551 & 0.2735 & 20.95 & 0.8324 \\
+ \ours & 0.3861 & 0.1475 & 0.2557 & 0.2741 & 21.93 & 0.8402 \\
ENS-MCMark & 0.3842 & 0.1459 & 0.2432 & 0.2725 & 21.08 & 0.8325 \\
+ \ours & 0.3845 & 0.1466 & 0.2432 & 0.2730 & 22.00 & 0.8399 \\ \bottomrule
\end{tabular}
\end{table}

\subsection{Extension to Longer Blocks}\label{sec:analysis-three-token}

\ours\ naturally extends from two-token blocks to longer token blocks.
In principle, watermarking \(r\)-token blocks further increases the effective vocabulary from \(V\) to \(V^r\) and can expose additional conditional entropy.
Table~\ref{tab:three-token-extension} reports results for a three-token variant on C4 with Llama3.2-3B-Instruct.
The three-token variant improves direct detection for all three base watermarks, particularly at short generation lengths and stringent false-positive rates.
For example, at 200 tokens and FPR \(=0.001\%\), the three-token variant improves over the two-token variant from \(47.2\%\) to \(63.0\%\) for ENS-DiPmark, from \(73.2\%\) to \(82.6\%\) for SynthID-Text, and from \(81.8\%\) to \(87.0\%\) for ENS-MCMark.

At the same time, longer blocks require more conditional distributions and are more sensitive to boundary disruption under editing attacks.
We therefore use two-token blocks as the default design point, which captures a substantial fraction of the available gain while preserving efficient generation and simple detection.

\begin{table}[t]
\centering
\caption{Direct detection when extending \ours{} from two-token blocks to three-token blocks on C4 using Llama3.2-3B-Instruct. Longer blocks further improve detection, especially at stringent false-positive rates.}
\label{tab:three-token-extension}
\setlength{\tabcolsep}{2pt}
\begin{tabular}{@{}l|cccc|cccc@{}}
\toprule
 & \multicolumn{4}{c|}{200 Tokens} & \multicolumn{4}{c}{400 Tokens} \\ \midrule
 & \multicolumn{3}{c|}{TPR@FPR=} & \multirow{2}{*}{\begin{tabular}[c]{@{}c@{}}Median \\ $p$-value $\downarrow$\end{tabular}} & \multicolumn{3}{c|}{TPR@FPR=} & \multirow{2}{*}{\begin{tabular}[c]{@{}c@{}}Median \\ $p$-value $\downarrow$\end{tabular}} \\ \cmidrule(lr){2-4} \cmidrule(lr){6-8}
 & 0.1\% $\uparrow$ & 0.01\% $\uparrow$ & \multicolumn{1}{c|}{0.001\% $\uparrow$} &  & 0.1\% $\uparrow$ & 0.01\% $\uparrow$ & \multicolumn{1}{c|}{0.001\% $\uparrow$} &  \\ \midrule
ENS-DiPmark & 47.0\% & 32.8\% & \multicolumn{1}{c|}{23.9\%} & 1.89e-3 & 73.8\% & 61.2\% & \multicolumn{1}{c|}{46.8\%} & 1.63e-5 \\
+ \ours\ 2-Token & 68.3\% & 58.9\% & \multicolumn{1}{c|}{47.2\%} & 2.01e-5 & 94.5\% & 88.6\% & \multicolumn{1}{c|}{81.7\%} & 4.03e-10 \\
+ \ours\ 3-Token & \textbf{77.0\%} & \textbf{70.8\%} & \multicolumn{1}{c|}{\textbf{63.0\%}} & \textbf{1.43e-7} & \textbf{98.7\%} & \textbf{96.2\%} & \multicolumn{1}{c|}{\textbf{93.0\%}} & \textbf{6.42e-14} \\ \midrule
SynthID-Text & 71.8\% & 60.5\% & \multicolumn{1}{c|}{50.3\%} & 9.59e-6 & 92.9\% & 86.5\% & \multicolumn{1}{c|}{80.1\%} & 1.73e-9 \\
+ \ours\ 2-Token & 87.5\% & 79.5\% & \multicolumn{1}{c|}{73.2\%} & 5.79e-10 & 99.3\% & 98.5\% & \multicolumn{1}{c|}{97.7\%} & 8.25e-18 \\
+ \ours\ 3-Token & \textbf{92.8\%} & \textbf{87.7\%} & \multicolumn{1}{c|}{\textbf{82.6\%}} & \textbf{3.81e-14} & \textbf{99.9\%} & \textbf{99.6\%} & \multicolumn{1}{c|}{\textbf{98.6\%}} & \textbf{6.84e-26} \\ \midrule
ENS-MCMark & 73.4\% & 62.5\% & \multicolumn{1}{c|}{54.0\%} & 3.86e-6 & 95.4\% & 91.3\% & \multicolumn{1}{c|}{83.5\%} & 2.00e-10 \\
+ \ours\ 2-Token & 92.0\% & 85.7\% & \multicolumn{1}{c|}{81.8\%} & 1.84e-11 & 99.6\% & 99.0\% & \multicolumn{1}{c|}{98.5\%} & 9.36e-21 \\
+ \ours\ 3-Token & \textbf{93.4\%} & \textbf{90.5\%} & \multicolumn{1}{c|}{\textbf{87.0\%}} & \textbf{9.08e-15} & \textbf{99.8\%} & \textbf{99.5\%} & \multicolumn{1}{c|}{\textbf{99.5\%}} & \textbf{1.16e-27} \\ \bottomrule
\end{tabular}
\end{table}

\end{document}